\documentclass{IEEEcsmag}

\usepackage[utf8]{inputenc}
\usepackage{lmodern}
\usepackage[T1]{fontenc}

\PassOptionsToPackage{hyphens}{url}
\usepackage[colorlinks,urlcolor=blue,linkcolor=blue,citecolor=blue]{hyperref}
\expandafter\def\expandafter\UrlBreaks\expandafter{\UrlBreaks\do\/\do\*\do\-\do\~\do\'\do\"\do\-}
\usepackage{upmath,color}

\usepackage{color,soul}

\jvol{XX}
\jnum{XX}
\pubyear{2024}

\usepackage[most]{tcolorbox}

\definecolor{usercolor}{RGB}{173, 216, 230} 
\definecolor{assistantcolor}{RGB}{255, 228, 196} 

\tcbset{
    userbox/.style={
        colback=usercolor,
        colframe=usercolor!50!black,
        fonttitle=\bfseries,
        title=User Prompt:
    },
    assistantbox/.style={
        colback=assistantcolor,
        colframe=assistantcolor!50!black,
        fonttitle=\bfseries,
        title=LLM Assistant:
    }
}

\newcommand{\user}[1]{\begin{tcolorbox}[userbox] #1 \end{tcolorbox}}
\newcommand{\assistant}[1]{\begin{tcolorbox}[assistantbox] #1 \end{tcolorbox}}

\begin{document}

\title{From Specifications to Prompts: On the Future of Generative LLMs in Requirements Engineering}

\author{Andreas Vogelsang}
\affil{University of Cologne, Germany}


\begin{abstract}
Generative LLMs, such as GPT, have the potential to revolutionize Requirements Engineering (RE) by automating tasks in new ways. This column explores the novelties and introduces the importance of precise prompts for effective interactions. Human evaluation and prompt engineering are essential in leveraging LLM capabilities.
\end{abstract}

\maketitle

\section{FROM THE EDITOR}
The theme of this issue is Generative AI for Software Engineering, a versatile approach that will impact the entire development lifecycle. Frequent readers will recall the summary of a panel discussion at the RE 2023 Conference from three issues ago. In this guest column, Andreas Vogelsang, drawing from his keynote at the 7th Workshop on Natural Language Processing for RE, continues this discussion. He explores the potential of generative AI to automate tasks, from capturing requirements to verifying consistency and completeness. Additionally, he discusses how RE can enhance prompt engineering. I'm sure generative AI will quickly evolve from a novel tool to an integral part of the software engineering context. Here are some insights from an expert at the forefront of this shift!

\section{Introduction}
Generative LLMs, such as GPT, have revolutionized our interaction with artificial intelligence. Their ability to understand, generate, and manipulate language presents unprecedented opportunities and challenges across various disciplines, including RE. Generative LLMs have the potential to redefine the landscape of requirements elicitation, specification, and validation.
%
%
%
The main messages conveyed in this column are:
\begin{enumerate}
    \item Generative LLMs provide novel features to support RE tasks. However, the human in the loop becomes even more important for evaluating generative LLMs. 
    \item Prompts are expressions of requirements. RE can help improve prompt engineering and position prompts as artifacts in the software development lifecycle.
\end{enumerate}

\section{Preliminaries: Decoder-only (Generative) LLMs}
Generative LLMs build upon a decoder-only architecture. Decoder-only architectures are derived from the well-known transformer architecture~\cite{Vaswani17}. The original transformer architecture addressed translation tasks by combining an encoder, which encodes the input, and a decoder, which generates the output. However, it turned out that the encoder and the decoder parts alone can be adapted to create helpful models. For example, the successful BERT model~\cite{Devlin18} is an encoder-only model. Encoder-only models are primarily used for predictive tasks where patterns in data are used to make forecasts and predictions (e.g., classification, clustering, similarity estimation).

Decoder-only LLMs have been designed to generate text. To support the generative capabilities of decoder-only LLMs, they are primarily pre-trained with a next-word prediction (NWP) objective, where the models predict the next word or words in a given sequence of words. After pre-training, decoder-only LLMs are triggered by a so-called \textit{prompt}. A prompt is a textual input instructing the generative LLM to generate the desired response.

Feeding decoder-only LLMs with prompts offers a new paradigm for interaction. In encoder-only or full transformer models, information about the task had to be reflected in the training data. The input had to be preprocessed before the model could make a prediction.
In decoder-only models, the task and the input can be expressed in natural language and passed directly to the model. The model's output is also richer than for non-generative LLMs since it produces (customizable) text instead of confidence values for predefined outcomes. A prompt contains any of the following elements:
\begin{itemize}
    \item \textbf{Instruction:} a specific task or instruction you want the model to perform
    \item \textbf{Context:} external information or additional context that can steer the model to better responses
    \item \textbf{Input Data:} the input or question that we are interested in finding a response for
    \item \textbf{Output Indicator:} the type or format of the output.
\end{itemize}
You do not need all four elements for a prompt, and the ideal format depends on the task. Although prompting an LLM sounds relatively straightforward, the creation and exact phrasing of a prompt is crucial for the quality of the LLM output. For most tasks, it is necessary to experiment with different prompts and iteratively refine them to yield the best results. This so-called \textit{prompt engineering} step is similar to feature engineering in more traditional machine learning (ML) approaches. Prompt engineering is a relatively new field where only heuristics and simple rules exist that help create effective prompts. 

\section{Using Generative LLMs for RE tasks}

Generative LLMs provide a new way to solve RE tasks. Traditional predictive LLMs (i.e., encoder-only LLMs) are trained or fine-tuned to predict the most likely output for a given input. Usually, the set of possible outputs is constrained. It is either a set of classes~\cite{winkler_automatic_2016, binder_automatically_2023}, a subset of the input (i.e., information retrieval~\cite{fischbach_towards_2020}), or a numeric value in a defined range (e.g., similarity estimation).  

With generative LLMs, a specific RE task is written as a prompt, which serves as input. The prompt usually contains the task and the input data to be processed. The model's output provides the answer to the specific task. 

Here is an example of how a trace link recovery task can be formulated as prompt~\cite{Rodriguez23}:

\user{Below are artifacts from the same software system. Is there a traceability link between (1) and (2)?
\vspace{1ex}

(1) 
The DPU-TMALI shall utilize SCM\_DCI\_SR, along with ERRNO provided by DPU-DCI to decode errors and place them on an error queue for DPU-CCM.
\vspace{1ex}

(2) 
Error Collection and Reporting The ccmErrEnq() function tracks the last error reported and its frequency of occurrence. Once an error code has been reported it becomes the previously reported error code maintained by ccmErrEnq(). A repetition count is then incremented for each subsequent, consecutively reported, identical instance of this previously reported error. If this error code is reported more than once in one high-rate housekeeping reporting period, then a special error, S\_ccm\_ERR\_REPEAT is enqueued with the repetition count for the error encoded in the least significant byte. This mechanism effectively reduces the potential for housekeeping telemetry to become flooded with a single repeated error.
}

\assistant{
Yes, there is a traceability link between (1) and (2). 
(1) mentions that the DPU-TMALI places errors on an error queue for the DPU-CCM.
\vspace{1ex}

(2) describes the error collection and reporting mechanism of the DPU-CCM, specifically referring to enqueuing errors.
\vspace{1ex}

So (2) provides an implementation detail for the error queuing functionality briefly mentioned in (1). Therefore, there is a traceability link between these two artifacts in the software system.
}

The quality of the provided output depends on the specific prompt. Prompts can be refined by refining any component (i.e., instructions, context, input question). 
Providing more context and instructions is an effective strategy to increase the semantic quality of the output. In a recent publication, we describe \textit{few-shot prompting}, \textit{instruction fine-tuning}, and \textit{retrieval augmented generation}, which can be used to refine the prompt and increase the model's performance~\cite{vogelsang_using_2024}.

Besides the novel input paradigm, LLMs also offer two novel ways to work with the output.

\textbf{We can ask generative LLMs to augment the response with rationales and explanations.}
In contrast to predictive LLMs, which only provide the prediction as output, generative LLMs can be instructed to provide explanations along the result. As also shown in the example above, the LLM assistant provides the answer (``yes, there is a trace link'') followed by an explanation. This additional information impacts the perception of the human analyst using the tool. It may also increase the chances that humans can uncover mispredictions by the model and also help humans to improve quality flaws. 

\textbf{We can enter into a dialog with a generative LLM.}
The second novel advantage is that generative LLMs can incorporate prior answers as context into new prompts. This gives the user the possibility to ask for additional information or even correct certain details of an answer. The generative LLM can respond to this additional human input to refine, clarify, or even correct its answer. 
If, for example, we have spotted a mistake in the answer, we may correct the model and get a refined answer:

\user{The explanation is incorrect. (2) does not describe the DPU-CCM but another component.}

\assistant{I apologize for the confusion in my previous response. Based on the information provided, there is no direct traceability link between (1) and (2) in terms of specific components or functionalities. (1) describes the error handling process in the DPU-TMALI component, while (2) describes the error collection and reporting mechanism in a different component that is not explicitly mentioned.

Without further context or explicit connections between the components mentioned in (1) and (2), it is not possible to establish a direct traceability link between the two artifacts.}

\textbf{We can combine different modalities.}
Although LLMs were originally designed to process and produce text, recent LLMs can also process and generate image, audio, and even video data. Much requirements-related information is in non-text format (e.g., visual models, interview recordings, vision videos). The recent LLM advancements allow the integration of these representations into the automation pipeline.

\textbf{Evaluation becomes a challenge.}
Augment a response with explanations, entering into a dialog, and combining modalities are features of generative LLMs that distinguish them from predictive LLMs we used in the past to automate RE tasks. These features, however, make the evaluation of LLMs more challenging.
When generative LLMs are used like predictive models, the evaluation is straightforward and should adhere to existing best practices (e.g., as suggested by Dell'Anna~et~al.~\cite{ECSER}).
However, when using the generative, conversational, and multimodal capabilities, the human-in-the-loop becomes much more important for evaluating generative LLMs.  
This starts with different metrics that are used to compare generated text (e.g., BLEU and ROUGE instead of precision and recall). Even more importantly, the performance of an LLM assistant will strongly depend on the conversational flow, the perceived quality of answers, the trust of humans in the answers, and the way the assistant handles human feedback.

\section{Using RE for Effective Prompting}

In the rapidly advancing landscape of AI, prompt engineering emerged as a new discipline concerning the development and optimization of prompts for LLMs. Various prompting techniques have been introduced by different sources, ranging from blog articles to peer-reviewed academic publications. However, the vast majority seem to lack a scientific foundation and are more based on trial and error. 
Some of these rules address the style of how prompts should be phrased, e.g.,

\begin{itemize}
    \item \textbf{Start simple:} start with a simple prompt and build on it
    \item \textbf{Call to action:} start the prompt with an action word like ``Write'', ``Create'', or ``Summarize'' instead of ``Can you''
    \item \textbf{Add context:} add specific and relevant context to the task you want to perform
    \item \textbf{Add expectations:} add clear and direct expectations for the content, like how long it should be and what to include
\end{itemize}

Prompt templates such as the CRISPE framework\footnote{\url{https://sourcingdenis.medium.com/crispe-prompt-engineering-framework-e47eaaf83611}} (\textbf{C}apacity and \textbf{R}ole, \textbf{I}nsight, \textbf{S}tatement, \textbf{P}ersonality, \textbf{E}xperiment) or the RICE framework\footnote{\url{https://www.geeky-gadgets.com/harnessing-the-power-of-the-rice-framework-for-perfect-chatgpt-prompts/}} (\textbf{R}ole, \textbf{I}nstructions, \textbf{C}ontext, \textbf{C}onstraints, and \textbf{E}xamples) provide guidelines and best practices for structuring the content of a prompt. A prompt for a trace link recovery task following the CRISPE framework could look like this:

\user{
\textbf{Capacity and Role:} Requirement Analyst

\textbf{Insight:} We're looking for traceability links between high-level and low-level requirements. A traceability link indicates that a low-level requirement is related to a high-level requirement (e.g. because it refines it).

\textbf{Statement:} Find all traceability links between the following low-level and high-level requirements.

\textbf{Personality:} Accurate and precise.

\textbf{Experiment:} Provide just one answer.
\vspace{1ex}

[Low-level Requirements]:
\ldots

[High-level Requirements]:
\ldots
}




Other prompting strategies suggest adding examples (few-shot prompting) or breaking down the problem into smaller tasks (chain-of-thought prompting).

From an RE perspective, prompts express the user's needs and expectations to an LLM. Therefore, prompts are expressions of requirements. 
RE research has a long tradition of thinking about and coming up with methods to describe stakeholder needs in a precise manner. This also includes thinking about the role and importance of prompts in a broader development context and not just aiming at the quality of the resulting outputs. 
Generally, the goals of RE are (1) knowing all relevant requirements, (2) achieving a consensus among the stakeholders about these requirements, (3) documenting requirements appropriately, and (4) managing requirements systematically. 
Adapting these goals to requirements expressed as prompts leads to interesting questions: 

\noindent \textbf{Knowing all \textit{prompts}}
\begin{itemize}
    \item Have we included all relevant content in a prompt?
    \item Do we need additional prompts to improve the results?
\end{itemize}

\noindent \textbf{Achieving a consensus among the stakeholders about these \textit{prompts}:}
\begin{itemize}
    \item Do all stakeholders agree with the prompt?
    \item Are the perspectives of all stakeholders reflected in a (set of) prompts?
\end{itemize}

\noindent \textbf{Documenting \textit{prompts} appropriately}
\begin{itemize}
    \item Is there a need to document and store the prompts?
    \item Who may need to read, change, or reuse the prompts?
\end{itemize}

\noindent\textbf{Managing \textit{prompts} systematically}
\begin{itemize}
    \item Which (version of a) prompt generated which artifact?
    \item Which LLM (version) has been used?
    \item Which prompts have been derived from which requirements?
    \item Is all prompt content backed up by requirements?
\end{itemize}

Recently, we investigated reproducible indicators within prompts that may predict a loss of quality or flaw. We base these flaw indicators on established requirements smells, which are reliable indicators for requirements quality~\cite{Femmer2017, Frattini2022}. Our initial experiments showed interesting results. For example, in a code generation task, ChatGPT interpreted the vague requirement ``If the points of one player are too low, the game ends and the player loses the game.'' by setting the boundary to 0 points without asking for clarification or hinting at the ambiguity:

\user{We want to develop a game for mobile devices. The game has two players who throw a dice in each turn. Both players start with 10 points. If the dice shows an even number, the number is added to the points of the player, if it shows an uneven number, the points are subtracted. \textbf{If the points of one player are too low, the game ends and the player loses the game.} 
\vspace{1ex}

Create Python code that implements this game.
}

\assistant{Certainly! Below is a Python implementation of the game:

[\ldots]

if \textit{player1\_points <= 0}:

\hspace{1cm} print(``Player 1 has lost the game!'')

\hspace{1cm} break

[\ldots]
}

Considering prompts as a new manifestation of requirements is a novel and relevant research direction. This line of research should start with characterizing the role of prompts in the development process.
Should prompts be considered similarly to requirements expressed in requirements documents? Are they more like intermediate and transitory artifacts? Or should they be treated as very high-level code? 



\section{Conclusions}
Generative large language models (LLMs) are opening up new opportunities for requirements analysts, providing innovative tools to enhance their work. Unlike traditional predictive models, generative LLMs offer detailed explanations and allow users to engage in a dialogue to refine or correct responses. This brings exciting possibilities but also introduces challenges. We need to rethink how we evaluate these models' capabilities and develop new ways to assess their effectiveness, especially when humans are involved.

Additionally, there is a growing interest in the intersection of RE and prompt engineering. How can we create precise prompts or break down complex goals into actionable steps? Is there a need for systematic prompt documentation and management? This is an intriguing area for RE researchers and enthusiasts to explore. 

\bibliographystyle{IEEEtran}
\bibliography{references,my-refs}

\begin{thebibliography}{10}
\providecommand{\url}[1]{#1}
\csname url@samestyle\endcsname
\providecommand{\newblock}{\relax}
\providecommand{\bibinfo}[2]{#2}
\providecommand{\BIBentrySTDinterwordspacing}{\spaceskip=0pt\relax}
\providecommand{\BIBentryALTinterwordstretchfactor}{4}
\providecommand{\BIBentryALTinterwordspacing}{\spaceskip=\fontdimen2\font plus
\BIBentryALTinterwordstretchfactor\fontdimen3\font minus
  \fontdimen4\font\relax}
\providecommand{\BIBforeignlanguage}[2]{{%
\expandafter\ifx\csname l@#1\endcsname\relax
\typeout{** WARNING: IEEEtran.bst: No hyphenation pattern has been}%
\typeout{** loaded for the language `#1'. Using the pattern for}%
\typeout{** the default language instead.}%
\else
\language=\csname l@#1\endcsname
\fi
#2}}
\providecommand{\BIBdecl}{\relax}
\BIBdecl

\bibitem{Vaswani17}
A.~Vaswani, N.~Shazeer, N.~Parmar, J.~Uszkoreit, L.~Jones, A.~N. Gomez, L.~u.
  Kaiser, and I.~Polosukhin, ``Attention is all you need,'' in \emph{Advances
  in Neural Information Processing Systems}, I.~Guyon, U.~V. Luxburg,
  S.~Bengio, H.~Wallach, R.~Fergus, S.~Vishwanathan, and R.~Garnett, Eds.,
  vol.~30.\hskip 1em plus 0.5em minus 0.4em\relax Curran Associates, Inc.,
  2017.

\bibitem{Devlin18}
J.~Devlin, M.-W. Chang, K.~Lee, and K.~Toutanova, ``{BERT}: Pre-training of
  deep bidirectional transformers for language understanding,'' 2018.

\bibitem{winkler_automatic_2016}
J.~Winkler and A.~Vogelsang, ``Automatic classification of requirements based
  on convolutional neural networks,'' in \emph{2016 {IEEE} 24th {International}
  {Requirements} {Engineering} {Conference} {Workshops} ({REW})}.\hskip 1em
  plus 0.5em minus 0.4em\relax IEEE, 2016, pp. 39--45.

\bibitem{binder_automatically_2023}
M.~Binder, A.~Vogt, A.~Bajraktari, and A.~Vogelsang, ``Automatically
  {Classifying} {Kano} {Model} {Factors} in {App} {Reviews},'' in
  \emph{International {Working} {Conference} on {Requirements} {Engineering}:
  {Foundation} for {Software} {Quality}}.\hskip 1em plus 0.5em minus
  0.4em\relax Springer Nature Switzerland Cham, 2023, pp. 245--261.

\bibitem{fischbach_towards_2020}
J.~Fischbach, B.~Hauptmann, L.~Konwitschny, D.~Spies, and A.~Vogelsang,
  ``Towards causality extraction from requirements,'' in \emph{{IEEE} 28th
  {International} {Requirements} {Engineering} {Conference} ({RE})}.\hskip 1em
  plus 0.5em minus 0.4em\relax IEEE, 2020, pp. 388--393.

\bibitem{Rodriguez23}
A.~D. Rodriguez, K.~R. Dearstyne, and J.~Cleland-Huang, ``Prompts matter:
  Insights and strategies for prompt engineering in automated software
  traceability,'' in \emph{IEEE 31st International Requirements Engineering
  Conference Workshops (REW)}, 2023, pp. 455--464.

\bibitem{vogelsang_using_2024}
A.~Vogelsang and J.~Fischbach, ``Using {Large} {Language} {Models} for
  {Natural} {Language} {Processing} {Tasks} in {Requirements} {Engineering}:
  {A} {Systematic} {Guideline},'' in \emph{Handbook of {Natural} {Language}
  {Processing} for {Requirements} {Engineering}}, A.~Ferrari and G.~G.
  Deshpande, Eds.\hskip 1em plus 0.5em minus 0.4em\relax Cham: Springer
  International Publishing, 2024.

\bibitem{ECSER}
D.~Dell'Anna, F.~B. Aydemir, and F.~Dalpiaz, ``Evaluating classifiers in {SE}
  research: the {ECSER} pipeline and two replication studies,'' \emph{Empirical
  Software Engineering}, vol.~28, no.~1, Nov. 2022.

\bibitem{Femmer2017}
H.~Femmer, D.~Méndez~Fernández, S.~Wagner, and S.~Eder, ``Rapid quality
  assurance with requirements smells,'' \emph{Journal of Systems and Software},
  vol. 123, pp. 190--213, 2017.

\bibitem{Frattini2022}
J.~Frattini, L.~Montgomery, J.~Fischbach, M.~Unterkalmsteiner, D.~Mendez, and
  D.~Fucci, ``A live extensible ontology of quality factors for textual
  requirements,'' in \emph{IEEE 30th International Requirements Engineering
  Conference (RE)}.\hskip 1em plus 0.5em minus 0.4em\relax IEEE, 2022.

\end{thebibliography}
\vspace*{-8pt}

\begin{IEEEbiography}{Andreas Vogelsang}{\,} is a full professor of Software Engineering at the University of Cologne, 50923 Cologne, Germany.  
Contact him at vogelsang@cs.uni-koeln.de.\vspace*{8pt}
\end{IEEEbiography}

\end{document}